\documentclass[aps,showpacs,nofootinbib,superscriptaddress]{revtex4}

\usepackage{epsfig}
\usepackage{graphicx}

\begin{document}

\title{Renormalization of Singlet NN-Scattering with One Pion Exchange
       and Boundary Conditions}\author{M. Pav\'on
       Valderrama}\email{mpavon@ugr.es} \affiliation{Departamento
       de F\'{\i}sica Moderna, Universidad de Granada, E-18071
       Granada, Spain.}  \author{E. Ruiz
       Arriola}\email{earriola@ugr.es} \affiliation{Departamento de
       F\'{\i}sica Moderna, Universidad de Granada, E-18071 Granada,
       Spain.}

\date{\today}

\begin{abstract} 
\rule{0ex}{3ex} We present a simple and physically compelling boundary
condition regularization scheme in the framework of effective field
theory as applied to nucleon-nucleon interaction. It is free of
off-shell ambiguities and ultraviolet divergences and provides finite
results at any step of the calculation. Low energy constants and their
non-perturbative evolution can directly be obtained from experimental
threshold parameters in a completely unique and model independent way
when the long range explicit pion effects are removed. This allows to
compute scattering phase shifts which are, by construction consistent
with effective range expansion to a given order in the CM momentum and
are free from finite cut-off artifacts. We illustrate how the method
works in the $^1S_0$ channel for the One Pion Exchange potential.
\end{abstract}

\pacs{03.65.Nk,11.10.Gh,13.75.Cs,21.30.Fe,21.45.+v}

\maketitle



\section{Introduction}

Effective field theories (EFT) have been successfully investigated in
recent years in the context of hadronic and nuclear physics. Their
main ingredient has to do with the occurrence of scale separation
between long and short distance physics, making the development of a
systematic power counting possible. Since the original proposal of
Weinberg's~\cite{Weinberg:rz} to make a power counting in the
potential many works have followed implementing such a
counting~\cite{Ordonez:1995rz,Park:1997kp,Epelbaum:1998ka,Entem:2002sf}
with finite cut-offs or proposing a counting in the renormalized
S-matrix~\cite{Kaplan:1998we,Geg98} which has also been
pursued~\cite{Fleming:1999ee}. Both Weinberg and Kaplan-Savage-Wise
schemes can be understood as perturbative expansions about infrared
fixed points~\cite{Birse:1998dk} (see also
Ref.~\cite{Barford:2001sx}). In any case, converge improves under
certain conditions~\cite{Phillips:1999hh}.  According to
Ref.~\cite{Beane:2001bc} a hybrid counting involving also the chiral
limit should be invoked (see also Ref.~\cite{Oller:2002dc}).  For a
recent and more complete review on these and related issues see
e.g. Ref.~\cite{Bedaque:2002mn} and references therein.

Much theoretical insight has been gained by analysing how short and
long distance physics separate for the One Pion Exchange (OPE)
interaction in the singlet $^1S_0$ channel where the scattering
length, $\alpha_0 = -23.7 {\rm fm} $, is much larger than the size of
the potential $ 1/m_\pi = 1.4 {\rm fm}$. The renormalization of NN
interaction in this channel has been studied several times in the
literature. In Ref.~\cite{Frederico:1999ps} an elegant subtraction
method has been developed to construct a finite $T$ matrix for
contact, i.e. zero range, interactions added to OPE. Renormalization
is indeed achieved by taking the subtraction scale to be much larger
than any other mass scale and checking for independence of results in
this limit. The resulting description of the $^1S_0$ is only valid to
very low energies, requiring for inclusion of derivative
terms. Unfortunately, the method cannot be easily extended in that
case. These derivative interactions can be included within a cut-off
regularization~\cite{Gegelia:2001ev,Eiras:2001hu} and dimensional
regularization~\cite{Nieves:2003uu}. In this latter case a
three-parameter fit can be achieved with no explicit two pion exchange
contribution. For the pionless theory, though, the inconsistency
between both regularization methods after renormalization has been
pointed out~\cite{Phillips:1997tn}. Momentum space treatments based on
the Lippmann-Schwinger equation appear more natural from a
diagrammatic point of view within a Lagrangian framework and allow
explicit consideration of nonlocal potentials. In practice, however,
the long range potentials used in NN scattering are local, and for
those the analysis of renormalization in coordinate space may be
simpler. In addition, the Schr\"odinger equation is a second order
operator and boundary conditions define a complete solution of the
problem in the whole space both inside and outside the boundary.  This
is equivalent to a sharp separation between the interior and exterior
region.  This property is naturally formulated in coordinate space for
a local potential.

Although the idea of using boundary conditions for NN scattering is a
rather old one (see e.g. Ref.~\cite{LF67} and references therein),
there have been recent works in this regard motivated by the
developments within
EFT~\cite{Phillips:1996ae,Cohen:1998bv,vanKolck:1998bw}. Actually, it
has been shown~\cite{vanKolck:1998bw} that in the absence of long
range forces a low momentum expansion of the potential within EFT
framework for the Lippmann-Schwinger equation is completely equivalent
to an effective range expansion and also to an energy expansion of a
generic boundary condition at the origin in coordinate space for the
Schr\"odinger equation. If a long range OPE potential is added we will
show below that due to the short distance Coulomb nature of this
potential the origin must be reached continuously from above $R \to 0
\, , \, R > 0 $ (i.e. excluding the point $R=0$), in harmony with
known theorems on self-adjoint extensions of Schr\"odinger
operators~\cite{Albeverio}.

In this paper we analyze precisely how the energy dependent boundary
condition must change as we move the boundary radius for fixed energy
to achieve independence of physical observables such as scattering
phase shifts. By doing so we are effectively changing the Hilbert
space since the wave function in the outer region is defined only from
the boundary to infinity. An advantage of this procedure is that we
never need to invoke off-shellness explicitly; at any step
we are dealing with an on-shell problem. In addition, we work directly
with finite quantities and no divergences appear at any step of the
calculation when the boundary radius is taken to zero from above.

\section{Variable phase equation with Boundary Conditions} 

The reduced Schr\"odinger equation for including OPE in the $^1S_0$
channel for NN-scattering with CM momentum $k$ reads
\begin{equation}
-u_k '' (r) + U(r) u_k (r) = k^2 u_k (r)    \, , 
\label{eq:sch} 
\end{equation}
together with the asymptotic condition at infinity  
\begin{equation}
u_k (r) \to \sin (k r + \delta(k))  \, . 
\label{eq:bcinfty} 
\end{equation}
The OPE potential in the $^1S_0$ channel reads
\begin{eqnarray}
U (r) &=& - \frac{g_A^2 m_\pi^2 M_N }{16 \pi f_\pi^2} \frac{e^{-m_\pi
r}}r \, .
\label{eq:OPE} 
\end{eqnarray}
Where $M_N$ is the nucleon mass, $m_\pi$ the pion mass, $f_\pi $ the
pion weak decay constant and $g_A $ the nucleon axial coupling
constant. In the numerical calculations below we take $M_N =938.92 \,
{\rm MeV}$, $f_\pi =93 \, {\rm MeV} $ , $m_\pi =138 \, {\rm MeV} $ and
$ g_A=1.25 $.  Our unknowledge of the interaction below a certain
distance scale $R$ is parameterized in terms of a boundary condition
at the matching point $r=R$,
\begin{equation}
u_k' (R) - L(k,R) u_k (R)=0 \, .
\label{eq:bcR} 
\end{equation}
In general, this boundary condition depends both on the boundary
radius $R$ and the momentum $k$. The value of $R$ separates the whole
space into two disjoint regions, an outer region where we assume the
interaction to be given by OPE potential, and an inner region where
interaction is regarded as unknown.

The boundary condition at $R$, Eq.~(\ref{eq:bcR}) has a simple
physical interpretation. If we switch off the long range piece $U(r)$
above the scale $R$, then the phase shift due to the short distance
physics below the scale $R$ is given by
\begin{equation}
\frac{u_k'(R)}{u_k(R)} = L(k,R) = k \cot (k R + \delta (k , R )) \, .
\end{equation}
It is interesting to see what kind of equation satisfies the short
distance phase shift, $\delta (k,R)$, as we steadily move the boundary
radius $R$ for a fixed momentum $k$. Using Schr\"odinger's equation at
the boundary $r=R$ we get the variable phase equation,
\begin{equation}
\frac{ d \delta (k,R) }{dR} = -\frac1k U(R) \sin^2 (k R+ \delta(k,R))
\, . 
\label{eq:vf}
\end{equation}
The obvious condition at infinity must be satisfied  
\begin{equation}
\lim_{R \to \infty} \delta(k,R) = \delta(k) \, 
\end{equation}
Thus, Eq.~(\ref{eq:vf}) describes the evolution of the phase shift as
we go down to lower distances, assuming that {\it both} the long
distance potential and the physical phase shift are known. Regardless
of whether or not the potential we are considering is realistic at
very short distances~\footnote{Two Pion Exchange becomes comparable to
OPE at about the distance of $r=1.5 \, {\rm fm}$. So, any
extrapolation of Eq.~(\ref{eq:vf}) with OPE below $1.5 \, {\rm fm}$
should not be considered realistic.} one can extrapolate the long
distance potential to the origin and define the zero range
OPE--extrapolated phase shift
\begin{equation}
\delta_S (k) = \lim_{R \to 0^+} \delta(k,R) \, . 
\end{equation}
The precise manner how this limit is built depends on the OPE
potential, Eq.~(\ref{eq:OPE}), and will be analyzed below.
Eq.~(\ref{eq:vf}) is well known in potential scattering ( for a review
see e.g. Ref.~\cite{Calogero}), but it has always been used assuming
the trivial initial condition $ \delta_S (k) = \lim_{R \to 0}
\delta(k,R) =0$.

\section{Low energy expansion of the boundary condition} 

The former variable phase equation, Eq.~(\ref{eq:vf}) can be cast in a
more convenient form by defining the variable $K-$matrix,
\begin{equation} 
K(k,R) = k \cot \delta(k,R) \, , 
\end{equation} 
yielding 
\begin{eqnarray}
\frac{ d K(k,R)}{dR} = U(R) \left[ K(k,R) \frac{\sin k R}k + \cos k R
\right]^2 \, . 
\label{eq:vk}
\end{eqnarray}
At low energies, however, it can be conveniently parameterized as an
effective range expansion, which carries over to the variable phase
\begin{equation}
k \cot \delta (k,R) = -\frac1{\alpha_0 (R) } + \frac12 r_0 (R)  k^2 + 
v_2 (R) k^4 \cdots 
\end{equation}
one has 
\begin{eqnarray}
\frac{d \alpha_0}{dR} &=& U(R) \left( \alpha_0 -R \right)^2
 \label{eq:valpha} \\ \frac{ d r_0}{dR} &=& 2 U(R) R^2 \left( 1- \frac
 {R}{ \alpha_0 } \right) \left( \frac{r_0}R + \frac{R}{3 \alpha_0 } -1
 \right) \label{eq:vr0}\\ \frac{d v_2 }{d R} &=& R^4 U(R) \left\{
 \frac14 \left( \frac{r_0}R + \frac{R}{3 \alpha_0 } -1 \right)^2 +
 2\left( 1- \frac{R}{ \alpha_0} \right) \left(-\frac1{12} \frac{r_0}R
 + \frac{v_2}{R^3} - \frac1{120} \frac{R}{ \alpha_0} + \frac1{24}\right) \right\}
\label{eq:vv2}
\end{eqnarray}
These equations have to be supplemented with some initial conditions
$\alpha_0 (R_0) $, $r_0 (R_0)$ and $v_2 (R_0) $ at a given boundary
radius, $R_0$.  If we take the initial boundary radius, $R_0=0$ the
set of equations, (\ref{eq:valpha}), (\ref{eq:vr0}) and (\ref{eq:vv2})
express the evolution of the low energy parameters at short-distances
when the long distance potential is switched on. Conversely, if the
initial boundary radius is taken to infinity they offer a possibility
to determine the short-distance low energy parameters from the total
ones by downwards evolution in the variable $R$ when the long distance
potential is switched off. Notice the very appealing and natural
hierarchy in the previous equations; while the distance evolution of
the scattering length $\alpha_0$ is autonomous, the remaining low
energy parameters $r_0 $, $ v_2$ , etc. depend on the previous ones.
The relation to the renormalization group approach of
Ref.~\cite{Birse:1998dk,Barford:2001sx} will be discussed
elsewhere~\cite{Pavon03}. 


Before presenting the numerical results (\ref{eq:valpha}),
(\ref{eq:vr0}) and (\ref{eq:vv2}) we analyze first the short and long
distance behaviour. At short distances $R << 1/m_\pi $ the OPE
potential behaves like the Coulomb potential. Eq.~(\ref{eq:valpha})
can be easily solved in two cases, $ \alpha_0 << R $ and $\alpha_0 >>
R $. While in the first case we get
\begin{eqnarray}
\alpha_0(R) &=& \alpha_0(R_0) - \frac{g_A^2 m_\pi^2 M_N }{32 \pi
f_\pi^2}\left( R^2 -R_0^2 \right) \qquad , \quad \alpha_0 << R
\label{eq:asy_reg} 
\end{eqnarray}
in the second case one solution behaves as
\begin{eqnarray}
\alpha_0(R) &=&\frac{\alpha_0(R_0)}{1+\alpha_0(R_0) \frac{g_\pi^2 m_\pi^2
M_N }{16 \pi f_\pi^2} \log (R/R_0)} \to \frac{16 \pi f_\pi^2}{g_A^2
m_\pi^2 M_N } \frac1{\log (R/R_0)} \qquad , \quad \alpha_0 << R
\label{eq:asy} 
\end{eqnarray}
where $ R < R_0 << 1/m_\pi $. As we see, $\alpha_0(R)$ goes to zero
very slowly and with $\alpha_0'(R) \to -\infty $ at short distances,
which in momentum space corresponds to the ultraviolet limit. The
resemblance of Eq.~(\ref{eq:asy}) with asymptotic freedom is
striking. It is easy to see that the first case,
Eq.~(\ref{eq:asy_reg}), corresponds to selecting the regular solution
at the origin, whereas Eq.~(\ref{eq:asy}) is the generic case, which
always contains an admixture of the irregular solution. Obviously, the
regular case is exceptional and for that particular situation one can
integrate from the origin starting with the trivial initial condition
$\delta(k,0)=0$ up to infinity.  The result corresponds to a pure OPE
interaction, with no short-distance interactions. The important thing
to note here is that no matter what the initial value of $\alpha_0$
was at infinity (except for the exceptional case discussed before),
removing one-pion exchange goes into the same value at the origin, as
implied by Eq.~(\ref{eq:asy})). This also means that any small
deviation of the $\alpha_0(R_0)$ at small distances results in huge
variations at infinity. Thus, removing OPE results in a extreme fine
tuning of the low energy parameters at short distances.


We analyze now the long distance behavior. Clearly, when $ R >>
1/m_\pi$ we have $\alpha_0^\prime (R)=0$, Eq.~(\ref{eq:valpha}), and we
approach quickly the asymptotic value $\alpha_0 (\infty) $. For such
long distances we can always use perturbation theory to solve the
equations backwards. For scattering lengths which are natural,
i.e. $\alpha_0 \sim 1/m_\pi $ we may neglect $\alpha_0(R)$ with
respect to $R$ and get
\begin{eqnarray}
\alpha_0 (R) - \alpha_0 &=& -\int_R^\infty U(R) R^2 dR + \dots   
\label{eq:LDE_nat}
\end{eqnarray}
For unnatural scattering lengths, $\alpha_0 >> 1/m_\pi $ we make the
opposite approximation, and get
\begin{eqnarray}
\frac1{\alpha_0 (R)} - \frac1{\alpha_0} &=& -\int_R^\infty U(R) dR + \dots  
\label{eq:LDE_unnat}
\end{eqnarray}
The previous equations (\ref{eq:LDE_nat}) and (\ref{eq:LDE_unnat})
hold irrespectively of the strength of the potential, provided $R$ is
sufficiently large. Similar approximations for the remaining low
energy parameters will be discussed elsewhere~\cite{Pavon03}.

\begin{figure}[]
\begin{center}
\epsfig{figure=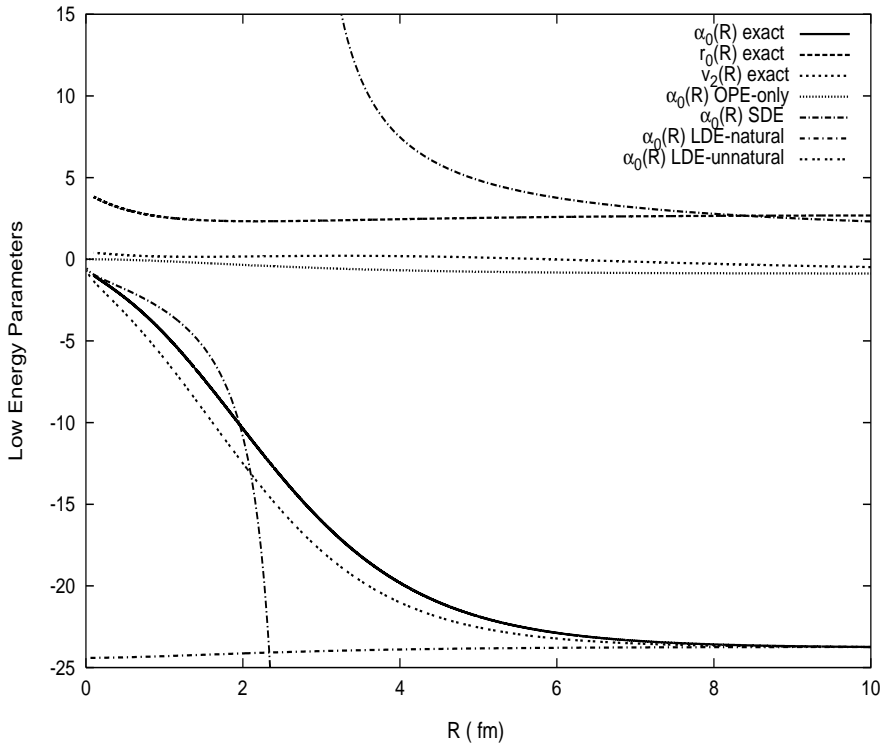,height=10cm,width=12cm}
\end{center}
\caption{Evolution of the scattering len
$^1S_0$ NN-threshold parameters $\alpha_0(R)$
(in $\, {\rm fm}$),$r_0(R)$ (in $\, {\rm fm}$) and $v_2(R)$ (in ${\rm
fm}^3$) from the asymptotic values at infinity (which we take in
practice $R_\infty= 10 \, {\rm fm}$ ) when OPE effects are removed
down to the origin. $\alpha_0 = -23.73 \, {\rm fm}$ and $r_0 = 2.68 \,
{\rm fm} $ and $v_2 = -0.48\, {\rm fm}^3 $. Solutions of
Eqs.~(\ref{eq:valpha},\ref{eq:vr0},\ref{eq:vv2}) are labelled as
``exact''. The extrapolated values at the origin when OPE effects are
removed are $ \alpha_{S,0} = 0 $, $ r_{S,0} = 4.08 \, {\rm fm}$ , and $
v_{S,2} = 0.43 \, {\rm fm}^3 $. We also show some approximations for
$\alpha_0 (R)$. OPE means One-Pion-Exchange only and corresponds to
integrate Eq.~(\ref{eq:valpha}) from the origin to infinity with the
boundary condition $\alpha_0(0)=0$. SDE means short distance expansion
as given by Eq.~(\ref{eq:asy}). LDE correspond to a long distance
expansion, Eq.~(\ref{eq:LDE_nat}) (natural case) and
Eq.~(\ref{eq:LDE_unnat}) (unnatural case), respectively. }
\label{fig:threshold_evol}
\end{figure}

The numerical evolution of $\alpha_{S,0} (R) $ and $r_S (R)$ starting
with the experimental values, $\alpha_0 = -23.739 \, {\rm fm}$, $r_0 =
2.68 \, {\rm fm} $ and $v_2 = -0.48 \, {\rm fm}^3 $ down to the origin
according to Eqs.~(\ref{eq:valpha},\ref{eq:vr0},\ref{eq:vv2}) is shown
in Fig.~(\ref{fig:threshold_evol})\footnote{In practice results are
insensitive for long distance cut-off of $ R_\infty= 10 {\rm fm} $. In
the case of the short distance cut-off we can go down to $R_S=0.0001
\, {\rm fm} $ without much effort but results are fairly insensitive
to the short distance radius already at $R_S = 0.1 \, {\rm fm}$, where
we have $\alpha_{S,0} = -0.9867 {\rm fm}$, $r_{0,S} = 3.819 {\rm fm} $
and $ v_{2,S} = 0.397 {\rm fm}^3 $. For shorter distances
Eq.~(\ref{eq:asy}) provides an accurate estimate for
$\alpha_0(R)$. Taking larger values of $R_s$ builds in finite cut-off
effects. Actually $R_S >> 1/m_\pi$ corresponds exactly to effective
range expansion.}.  We also show the perturbative estimate in the
case of large and small scattering lengths based on a long distance
expansion Eq.~(\ref{eq:LDE_nat}) (natural case) and
Eq.~(\ref{eq:LDE_unnat}) (unnatural case), respectively, as well as
our short distance estimate, Eq.~(\ref{eq:asy}). In the case of
$\alpha_0(R)$ we observe a huge change from infinity down to the
origin, although remains unnatural, $\alpha_0(R) >> R $. Numerically
we confirm our theoretical expectation that $\alpha_{S,0}(0)=0$ (see
Eq.~(\ref{eq:asy})).  This simply means that the renormalized contact
interaction becomes arbitrarily small as the OPE potential is switched
off. This is, however, not the case for the renormalized derivative
interaction, as expected from our estimate, Eq.~(\ref{eq:asy}).  Our
numerical values extrapolated to the origin are
\begin{eqnarray}
\alpha_{S,0} = \alpha_0 (0^+) &=& 0 \\ r_{S,0}= r_0 (0^+) &=& 4.08 {\rm
fm} \\ v_{S,2}=v_2 (0^+) &=& 0.43 \, {\rm fm}^3
\end{eqnarray}
This is the initial condition which, in principle, has to be
supplemented in Eqs.~(\ref{eq:valpha},\ref{eq:vr0},\ref{eq:vv2}) in
order to get the experimental results (see also discussion below). The
work of Ref.~\cite{Steele:1998zc} uses a two Yukawa model to extract
the short-distance low energy parameters. This is done by fitting the
data and then switching off the OPE contribution, yielding
$\alpha_{S,0} = -1.72 \, {\rm fm} $, $ r_{S,0}= 1.60 \, {\rm fm} $ and
$ v_{S,2}= -0.024 {\rm fm}^3 $. In Ref.~\cite{Beane:2001bc} an attempt
to determine the short-distance parameters based on the three Yukawa
model yields $\alpha_{S,0} = -3.38 \, {\rm fm} $, $ r_{S,0}= 2.60 \,
{\rm fm} $ and $ v_{S,2}= 0.313 \, {\rm fm}^3 $. The short-distance
scales in that calculation are $R_\rho= 2/m_\rho=0.46 \, {\rm fm} $
and $R_\sigma = 2/m_\sigma=0.80 \, {\rm fm} $. For that range we get
$\alpha_{S,0} = -3.6,-2.21 \, {\rm fm} $, $ r_{S,0}= 2.7,3.1 \, {\rm
fm} $ and $ v_{S,2}= 0.18,0.26 \, {\rm fm}^3 $ respectively, in
qualitative agreement with
Refs.~\cite{Steele:1998zc,Beane:2001bc}. Note, however that our way of
determining the short-distance low energy parameters does not require
any specific model at short distances.

\section{$^1S_0$-Phase shift} 

Once the short distance parameters are known one may compute the phase
shifts to any order of the approximation in a $k^2 $ expansion of the
initial condition {\it without any additional parameter fitting} by
integrating Eq.~(\ref{eq:vk}) upwards with a suitable initial
condition at a short distance radius, 
\begin{equation}
K_S(k)= k \cot \delta_S (k) = -\frac1{\alpha_{S,0}} + \frac12 r_{0,S} k^2
+ v_{2,S} k^4 + \cdots
\label{eq:eff_short}
\end{equation}
The standard way of proceeding is to determine the low energy
constants or equivalently the short distance parameters directly from
a fit to the data and then recompute the threshold parameters.  A
further advantage of avoiding a fit is that one can prevent spurious
and/or multiple minima; our solution is essentially unique. Morevover,
since by construction at a given order in the $k^2$ expansion the low
energy behavior of the phase shift is reproduced up to the same order
in $k^2$, the possibility of getting even slightly different threshold
parameters due to a fit in the intermediate energy region is
precluded.  Actually, our procedure would coincide with the standard
one, if the fit was carried out in the region where an effective range
expansion holds ($k < 60 {\rm MeV} $ if $v_2 $ is included).

Due to the fact that the origin is a fixed point for the running
scattering length, i.e.  $\alpha_0 (R) \to 0 $ for $R \to 0$ regardless
of the value of $\alpha_0(\infty)$, Eq.~(\ref{eq:asy}), one must
integrate the equations from very small distances upwards, using the
value of $\alpha_0 (R) $ at that distance. It is important to realize
that a tiny mismatch in the value of $\alpha_0$ close to the origin
results in a complete different value of $\alpha_0$ and also of the
phase shift at infinity

\begin{figure}[]
\begin{center}
\epsfig{figure=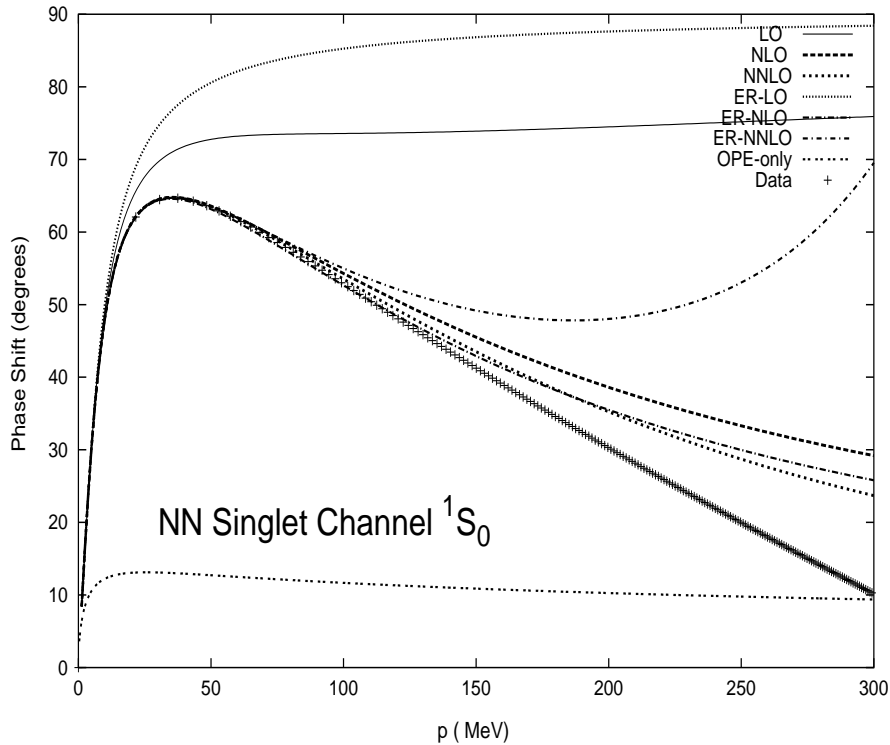,height=10cm,width=12cm}
\end{center}
\caption{Predicted phase shifts according to Eq.~(\ref{eq:vk}) when
OPE potential is switched on and the initial condition is a low energy
expansion of the $K$-matrix at short distances, (See
Eq.~(\ref{eq:eff_short}) in the main text). LO means keeping
$\alpha_{S,0}$ only, NLO keeping $ \alpha_{S,0}$ and $r_{0,S}$ and NNLO
keeping $ \alpha_{S,0}$, $r_{0,S}$ and $v_{2,S}$. The short range
parameters are directly determined by evolving the low energy
parameters from their experimental values $\alpha_0 = -23.73 \, {\rm
fm}$ and $r_0 = 2.68 \, {\rm fm} $ and $v_2 = -0.48 \, {\rm fm}^3
$. ER-LO, ER-NLO and ER-NNLO corresponds to a pure effective range
expansion keeping $\alpha_0$ only, $ \alpha_{S,0}$ and $r_{0}$, $ \alpha_0 $,
$r_0$ and $v_2$ respectively. OPE-only corresponds to OPE without
short-distance contributions.  No further fit is involved. Data are
the PWA from Ref.~\cite{Stoks:1993tb}.}
\label{fig:1S0_phase}
\end{figure}

In Fig.~(\ref{fig:1S0_phase}) we show the results for the phase shift
depending on the number of terms kept in the low energy expansion at
short distances.  Our results exhibit a good convergence rate.  For
comparison we also depict the effective range expansion results
without explicit pions, which is expected to work at low energies
only, and corresponds to make $R_S \to \infty$ in our approach. As we
see, the effect of introducing pions always improves the results. This
can be fully appreciated at NNLO, where ER does a poor job above CM
momenta $\sim 100 {\rm MeV}$, but explicit OPE effects enlarge the
energy range up to about $\sim 140 {\rm MeV} \sim m_\pi$.  where we
expect explicit two pion exchange contributions to start playing a
role.

An interesting point to note at this stage is that if $\alpha_{S,0} =
0 $ with other short-distance low energy parameters fixed, we would
inevitably get $\delta_S(k) = n \pi $, as deduced for instance from
Eq.~(\ref{eq:eff_short}). If we solve the variable phase equation with
that condition at $R=0$ up to $R=R_\infty >> 1/m_\pi $ we get the
result (also shown in Fig.~(\ref{fig:1S0_phase}) for comparison)
corresponding to a regular OPE with the regular boundary condition
$u_k (0)=0$ instead of the mixed boundary condition of
Eq.~(\ref{eq:bcR}) at $R=0$. The puzzle is resolved by realizing that
the limiting procedure in the boundary condition and the solution do
not commute; the limit $ R\to 0^+$ implies $\delta'(k,R) \to \infty $
whereas starting at $ R=0$ requires $\delta(k,R) \sim R^2 $ producing
instead a bound derivative $\delta'(k,R) \sim R $ (see
Eq.~(\ref{eq:vf})). This discontinuous dependence of the boundary
condition on the boundary radius at $R=0$ agrees with rigurous
theorems on self-adjoint extensions of Schr\"odinger operators (see
e. g. Appendix D of Ref.~\cite{Albeverio}).

\section{Conclusions} 

In the present paper we have analyzed the renormalization of the OPE
interaction in the presence of contact and derivative interactions of
any order for NN scattering. In order to do that we have derived an
equation for the evolution of an energy dependent boundary condition
in coordinate space as a function of the boundary radius. The
resulting equation shares many properties with renormalization group
equations and can be interpreted in terms of the phase shift produced
by eliminating OPE from infinity to the boundary radius, which
eventually is taken to zero. Two advantages can be deduced from this
framework: no divergences appear and there is no need to consider
off-shell extrapolations. This allows to set up equations for the
running low energy parameters as a function of the boundary
radius. Using the experimental values for the low energy parameters,
which correspond to an infinity boundary radius, we extract in a
unique and model independent way the corresponding short-distance
parameters. Our numerical values agree with other determinations based
on specific models for the short-distance interaction. As we get
closer to the origin we find a fixed point structure, triggered by the
non-vanishing contribution of the irregular solution. This requires a
fine tuning of the short-distance low energy parameters. After that we
integrate the running phase shift upwards and determine without any
additional fit the $^1S_0$ phase shift. The OPE plus contact and
derivative interactions to NNLO is able to describe the $^1S_0$ phase
shift up to C.M. momentum of about $140 \, {\rm MeV}$, which coincides
with opening of the two pion exchange left cut. Above that momentum
explicit two pion exchange effects should set in.

The results presented in this paper are very encouraging and suggest
several improvements and extensions. Explicit Two Pion Exchange
contributions are expected to contribute significantly at about $1.5-2
{\rm fm} $, so our results should not be considered realistic below
that scale, or equivalently above CM momenta of about $100-150 {\rm
MeV}$, as it seems to be the case. In addition, our description should
be enlarged to include all partial waves. Work along these lines will
be presented elsewhere~\cite{Pavon03}.

\begin{acknowledgments}

We thank M. C. Renmeester for useful correspondence and J. Nieves for
discussions and reading the manuscript. This research was supported by
DGI and FEDER funds, under contract BFM2002-03218 and by the Junta de
Andaluc\'\i a.

\end{acknowledgments}

\end{document}